\documentclass[cernpreprint,atlasdraft=false,texlive=2016,UKenglish,texmf]{atlasdoc}
 
\usepackage{atlaspackage}
\usepackage[backref=false]{atlasbiblatex}
 
\usepackage{atlasphysics}
 
\usepackage{multirow}
 
\graphicspath{{logos/}{figures/}}
 
\usepackage{ANA-HIGG-2019-14-PAPER-defs}

\AtlasTitle{A search for the dimuon decay of the Standard Model Higgs boson with the ATLAS detector}

\AtlasAbstract{
A search for the dimuon decay of the Standard Model (SM) Higgs boson is
performed using data corresponding to an integrated luminosity
of $139\,\mathrm{fb}^{-1}$ collected with the ATLAS detector in Run~2 $pp$
collisions at $\sqrt{s}=13\,$TeV at the Large Hadron Collider.
The observed (expected) significance over the background-only hypothesis
for a Higgs boson with a mass of $125.09\,$~GeV is $2.0\sigma$ ($1.7\sigma$).
The observed upper limit on the cross section times branching ratio for $pp \to H \to \mu\mu$ is
$2.2$ times the SM prediction at $95\%$ confidence level, while the expected limit on a $H \to \mu\mu$ signal
assuming the absence (presence) of a SM signal is $1.1$ ($2.0$).
The best-fit value of the signal strength parameter, defined as the ratio of the observed signal
yield to the one expected in the SM, is $\mu = 1.2 \pm 0.6$.}
 
\AtlasJournal{Phys. Lett. B}
\AtlasJournalRef{Phys. Lett. B 812 (2021) 135980}
\AtlasDOI{10.1016/j.physletb.2020.135980}
 
\author{The ATLAS Collaboration}
 
\AtlasRefCode{ANA-HIGG-2019-14}

\PreprintIdNumber{CERN-EP-2020-117}
\hypersetup{pdftitle={ATLAS document},pdfauthor={The ATLAS Collaboration}}
 
\begin{document}
 
\maketitle
 
\tableofcontents
 
\section{Introduction}
 
In July 2012, the ATLAS and CMS Collaborations announced the discovery
of a new particle with a mass of approximately
$125\GeV$~\cite{HIGG-2012-27, CMS-HIG-12-028} at the CERN Large Hadron
Collider (LHC). Subsequent measurements have indicated that this
particle is consistent with the Standard Model (SM) Higgs
boson~\cite{HIGG-2013-17, CMS-HIG-14-018, HIGG-2014-14, HIGG-2015-07},
denoted by~$H$.  While the interaction between the Higgs boson and the charged fermions
of the third-generation has already been observed by both the ATLAS and CMS
Collaborations~\cite{HIGG-2015-07, HIGG-2018-04, HIGG-2018-13, CMS-HIG-18-016, CMS-HIG-17-035},
only upper limits have been set on the interactions with fermions of the other generations.
The \hmm\ decay offers the best opportunity to
measure the Higgs interactions with a second-generation fermion at the LHC.
The SM branching ratio to dimuons for the Higgs boson with $m_H =
125.09\GeV$ is $(2.17 \pm 0.04) \times 10^{-4}$~\cite{deFlorian:2016spz}.
However, physics beyond the SM~\cite{Vignaroli:2009vt,Dery:2013rta} could
modify the branching ratio.
 
Both the ATLAS and CMS Collaborations carried out searches for the \hmm\ process based on partial
sets of about one quarter of the data collected during Run~2 of the LHC~\cite{HIGG-2016-10, CMS-HIG-17-019}.
This paper presents an improved search for the dimuon decay of the Higgs boson
using the full $pp$ collision dataset recorded with the ATLAS detector in the LHC Run~2 period, spanning 2015 to 2018
at $\sqrt{s}=13\TeV$, corresponding to an integrated luminosity of about $139\,\ifb$.
Compared to the previous publication~\cite{HIGG-2016-10}, several improvements
have been made. They include a better categorisation based on multivariate techniques
that exploit the topological and kinematic differences between the different signal production modes and the background processes, improvements in the muon reconstruction, a large increase in the equivalent integrated luminosity
of the simulated background samples using a dedicated fast simulation, and an improved methodology for
the background modelling.
 
The analysis selects events with two opposite-charge muons and
classifies them into 20 mutually exclusive categories based on the event topology and multivariate
discriminants to increase the signal sensitivity.
After event categorisation, the signal yield is extracted by a simultaneous fit to the 20 dimuon mass
(\mmm) distributions in the range 110--160\,\GeV\ together with background normalisation and shape parameters,
exploiting the resonant behaviour of the Higgs boson signal.
The Higgs boson is assumed to have a mass of $m_H = 125.09\GeV$~\cite{HIGG-2014-14} for all results presented.
More recent measurements of the Higgs boson mass~\cite{HIGG-2016-33,CMS-HIG-19-004} are compatible
within their uncertainties with this value.
 
 
\section{ATLAS detector}
 
The ATLAS detector \cite{PERF-2007-01,ATLAS-TDR-19} covers nearly the
entire solid angle around the collision point.\footnote{ATLAS uses a
right-handed coordinate system with its origin at the nominal
interaction point (IP) in the centre of the detector and the
$z$-axis along the beam pipe. The $x$-axis points from the IP to the
centre of the LHC ring, and the $y$-axis points upward. Cylindrical
coordinates $(r,\phi)$ are used in the transverse plane, $\phi$
being the azimuthal angle around the $z$-axis. The pseudorapidity is
defined in terms of the polar angle $\theta$ as
$\eta=-\ln\tan(\theta/2)$.  The rapidity is defined as
$y=\frac{1}{2}\ln\frac{E+p_z}{E-p_z}$ and the distance between two
objects is defined as $\Delta R = \sqrt{(\Delta y)^2 +
(\Delta\phi)^2}$. } It consists of an inner tracking detector
surrounded by a thin superconducting solenoid, electromagnetic and
hadronic calorimeters, and a muon spectrometer incorporating three
large superconducting toroid magnets.
 
The inner detector (ID) system is immersed in a 2~T axial magnetic
field and provides charged-particle tracking in the range $|\eta| <
2.5$. A high-granularity silicon pixel detector covers the vertex
region and typically provides four measurements per track. It is
surrounded by a silicon microstrip tracker, which typically provides four
measurement points per track. These silicon detectors
are complemented by a transition radiation tracker, which enables
radially extended track reconstruction up to $|\eta| = 2.0$.
 
The calorimeter system covers the pseudorapidity range $|\eta| <
4.9$. Within the region $|\eta| < 3.2$, electromagnetic calorimetry is
provided by barrel and endcap high-granularity liquid-argon (LAr)
sampling calorimeters, with an additional thin LAr presampler
covering $|\eta| < 1.8$ to correct for energy loss in material
upstream of the calorimeters. Hadronic calorimetry is provided by a
scintillator-tile calorimeter, segmented into three barrel structures
within $|\eta| < 1.7$, and two LAr hadronic endcap calorimeters.
 
The muon spectrometer (MS) comprises separate trigger and
high-precision tracking chambers measuring the deflection of muons in
a magnetic field generated by the superconducting air-core toroids. The
precision chamber system covers the region $|\eta| < 2.7$ with three
layers of monitored drift tubes, complemented by cathode-strip
chambers in the forward region, where the background is highest. The
muon trigger system covers the range $|\eta| < 2.4$ with
resistive-plate chambers in the barrel, and thin-gap chambers in the
endcap regions.
 
The data were collected with a two-level trigger
system~\cite{TRIG-2016-01}. The first-level trigger (L1) is
implemented in hardware and uses a subset of the detector
information. This is followed by a software-based high-level trigger
which runs algorithms similar to those in the offline reconstruction
software, reducing the event rate to approximately 1~kHz from the
maximum L1 rate of 100~kHz.
 
\section{Data and simulated event samples}
\label{sec:samples}
 
The $pp$ collision data at $\sqrt{s} = 13\TeV$ analysed here
correspond to the full recorded Run~2 dataset, with an integrated luminosity of
\lumi\ after the application of data quality requirements.
The mean number of $pp$ interactions per bunch crossing was about 34.
Events used in this analysis were recorded using a combination of
single-muon triggers with transverse momentum thresholds up to
$26\GeV$ for isolated muons and $50\GeV$ for muons without any
isolation requirement imposed, allowing to recover some inefficiency introduced by the isolation requirement at trigger level for high momentum muons.
The trigger efficiency for the sum of the \hmm
signal processes is about $91\%$ relative to the common event preselection
discussed in \Sec~\ref{sec:event}.
 
Samples of simulated Monte Carlo (MC) events are used to optimise the
selection, to model the signal processes and to develop an analytic
function to model the \mmm\ distributions for the background
estimate. The signal samples as well as a complete set of background
processes were processed through the
full ATLAS detector simulation~\cite{SOFT-2010-01} based on
$\GEANT4$~\cite{Agostinelli:2002hh}, henceforth referred to as fully
simulated samples.
 
Signal samples were generated for the main Higgs boson production
modes. The mass of the Higgs boson was set in the
simulation to $m_H = 125 \GeV$ and the corresponding width is $\Gamma_H =
4.07\MeV$~\cite{Heinemeyer:2013tqa}. The samples are
normalised with the latest available theoretical calculations of the
corresponding SM production cross sections, summarised in
Ref.~\cite{deFlorian:2016spz}. The normalisation of all Higgs boson
samples also accounts for the \hmm\ branching ratio of $2.17 \times
10^{-4}$ calculated with HDECAY~\cite{Djouadi:1997yw,Spira:1997dg,Djouadi:2006bz,Djouadi_2019}
and PROPHECY4F~\cite{Bredenstein:2006ha,Bredenstein:2006rh,Bredenstein:2006nk}.
 
Higgs boson production in the gluon--gluon fusion process (\ggF) was simulated using the
\POWHEG NNLOPS program~\cite{Nason:2004rx,Frixione:2007vw,Alioli:2010xd,Campbell:2012am,Hamilton:2012np,Hamilton:2012rf,Hamilton:2013fea,Hamilton:2015nsa} with
the PDF4LHC15 set of parton distribution functions
(PDFs)~\cite{Butterworth:2015oua}. The simulation achieves
next-to-next-to-leading-order (NNLO) accuracy in QCD for
inclusive observables after reweighting the Higgs boson
rapidity spectrum~\cite{Catani:2007vq}.
The parton-level events were processed by
\PYTHIAV{8}~\cite{Sjostrand:2014zea} to decay the Higgs bosons and to provide parton showering, final-state photon radiation (QED FSR),
hadronisation and the underlying event, using the AZNLO set of tuned
parameters~\cite{STDM-2012-23}. The
sample is normalised to a next-to-next-to-next-to-leading-order QCD
calculation with  next-to-leading-order (NLO) electroweak corrections~\cite{Aglietti:2004nj,Actis:2008ug,Actis:2008ts,Anastasiou:2008tj,Pak:2009dg,Harlander:2009bw,Harlander:2009mq,Harlander:2009my,Actis:2008ug,Anastasiou:2015ema,Anastasiou:2016cez,Dulat:2018rbf,Bonetti:2018ukf}.
 
Higgs bosons produced via vector-boson fusion (VBF) and in association with a vector boson, $q\bar{q}/qg \to VH$ with $V=W$ or $Z$, were generated at NLO accuracy in QCD
using the \POWHEGBOX program~\cite{Nason:2009ai,Cullen:2011ac,Luisoni:2013kna}.
The loop-induced process $gg \to ZH$ was generated
at leading order (LO) using \POWHEGBOX. For the VBF and $VH$ samples the
same settings for the PDF set and \PYTHIAV{8} as in the \ggF\ sample were
employed. The VBF sample is normalised to an approximate-NNLO QCD
cross section with NLO electroweak corrections~\cite{Ciccolini:2007jr,Ciccolini:2007ec,Bolzoni:2010xr}. The $VH$ samples are normalised to cross sections calculated at NNLO in
QCD with NLO electroweak corrections for $q\bar{q}/qg \to VH$ and at NLO
and next-to-leading-logarithm accuracy in QCD for $gg \to
ZH$~\cite{Ciccolini:2003jy,Brein:2003wg,Brein:2011vx,Altenkamp:2012sx,Denner:2014cla,Brein:2012ne,Harlander:2014wda,Harlander:2018yio}.  Higgs boson
production in association with a top-quark pair, \ttH, was simulated at NLO accuracy in QCD using
\MGMCatNLO~\cite{Alwall:2014hca,Artoisenet:2012st} with the NNPDF3.0NLO PDF
set~\cite{Ball:2014uwa} and interfaced to \PYTHIAV{8} using the A14 set of tuned
parameters~\cite{ATL-PHYS-PUB-2014-021}. The cross section is
taken from a calculation accurate to NLO in QCD with NLO
electroweak corrections~\cite{Beenakker:2002nc,Dawson:2003zu,Yu:2014cka,Frixione:2015zaa}.
 
Background events from the Drell--Yan (DY) $Z/\gamma^* \to\mu\mu$ process
were generated with
\SHERPAV{2.2.1}~\cite{Bothmann:2019yzt} using NLO-accurate
matrix elements for up to two partons, and LO-accurate matrix elements
for up to four partons calculated with the Comix~\cite{Gleisberg:2008fv}
and OpenLoops~\cite{Cascioli:2011va,Denner:2016kdg} libraries and the
NNPDF3.0 NNLO set. They were matched to the
\SHERPA parton shower~\cite{Schumann:2007mg} using the MEPS@NLO
prescription~\cite{Hoeche:2011fd,Hoeche:2012yf,Catani:2001cc,Hoeche:2009rj}.
Diboson processes ($WW$, $WZ$, and $ZZ$) as well as electroweak $Zjj$
production were simulated in a similar set-up with \SHERPAV{2.2.1}. The
$t\bar{t}$ and single-top-quark samples were generated at NLO accuracy with
\POWHEGBOX~\cite{Frixione:2007nw,Re:2010bp} using the NNPDF3.0NLO PDF
set interfaced to \PYTHIAV{8} for parton showering and hadronisation using
the A14 parameter set. For the $Wt$ process, the diagram removal scheme~\cite{Frixione:2008yi} is applied to remove the overlap with $t\bar{t}$ production.
The production of \ttV events was modelled using
the \MGMCatNLO~\cite{Alwall:2014hca} generator at NLO in a set-up similar to the one used for the \ttH process.
 
The effects of multiple $pp$
collisions in the same or neighbouring bunch crossings (pile-up) are
included in the MC simulation by overlaying inelastic $pp$
interactions produced using \PYTHIAV{8} with the NNPDF2.3LO set of
PDFs~\cite{Ball:2012cx} and the A3 set of
tuned parameters~\cite{ATL-PHYS-PUB-2016-017}. Events are reweighted such that the
distribution of the average number of interactions per bunch crossing
matches that observed in data. Simulated events are corrected to
reflect the momentum scales and resolutions as well as
the trigger, reconstruction, identification, and isolation efficiencies measured
in data for all the physics objects used in this analysis.
 
The background samples discussed above provide an equivalent integrated luminosity
that is typically $5$--$20$ times higher than that of data and are used to train
multivariate classifiers and to test the background modelling.
However, the statistical uncertainties in the
dominant DY background are a limiting factor in studying the background modelling
at the level required by the small expected \hmm signal. Therefore, two
fast-simulation set-ups were developed to generate significantly larger DY samples.
The primary fast-simulation DY
sample is based on parton-level events generated~\cite{AcknoledgeHPCGrant} with \SHERPAV{2.2.4}~\cite{Hoeche:2019rti} using LO matrix elements for $Z/\gamma^*$
production with up to three additional partons and using the CT14 NNLO
PDF set~\cite{Dulat:2015mca}. The parton-level events were processed
with \PYTHIAV{8} to provide QED and QCD parton showering and hadronisation,
and double-counted QCD emissions were removed using the CKKW-L
algorithm~\cite{Lonnblad:2001iq} with a merging scale of $20\,\gev$.
For further cross-checks, an additional fast-simulation DY sample was
prepared that
simulates $Z/\gamma^*+0,1$ partons inclusively at NLO accuracy using
\POWHEGBOX~\cite{Barze:2013fru} with the CT10 PDF set~\cite{Lai:2010vv}
and $Z/\gamma^*+2$ partons at LO accuracy with
\ALPGEN~\cite{Mangano:2002ea} using the CTEQ6L1 PDF
set~\cite{Pumplin:2002vw}. These parton-level events were processed
with an approximate QCD shower algorithm, overlaps between the two samples
were removed, and QED
FSR was provided by \textsc{Photos}~\cite{Golonka:2006tw}. For both of these
generated samples, experimental effects were
approximated using parameterisations rather than using the full ATLAS detector simulation
and reconstruction software.
The parameterisations, extracted from fully simulated MC samples or directly from ATLAS data,
reproduce the reconstruction and selection
efficiencies of detector-level objects by event weighting and model the resolution of the ATLAS detector
with predetermined probability distributions. Detailed descriptions were employed
for the muon momentum resolution and muon trigger and selection efficiencies,
for the photons from QED FSR, for hadronic jets from the primary
interaction and pile-up events in terms of kinematics and the number
of associated ID tracks, and for the effect of pile-up and the
underlying event on the measurement of the missing transverse momentum
\met. In total, two sets of about $10$--$20$ billion events are prepared in this way,
corresponding to an equivalent integrated luminosity of at least $50\iab$ in the
kinematic phase space  relevant for the analysis.
 
Both fast simulation DY samples give a good description of the data distributions
of the observables used in this analysis to discriminate the DY background
from the \hmm signal, i.e. the \mmm mass spectra and the multivariate
discriminants described in the following sections. Small residual
differences are taken into account by reweighting the mass spectra to
the data sidebands as described in Section~\ref{sec:modelling}.
 
\section{Object definitions and event selection }
\label{sec:event}
 
Events are required to contain at least one reconstructed $pp$
collision vertex candidate with at least two associated ID tracks each
with $\pT > 0.5\GeV$.  The vertex with the largest sum of $\pT^2$ of
tracks is considered to be the primary vertex of the hard interaction. For signal events the primary vertex selection criteria has an efficiency of about 99\%~\cite{MUON-2018-03}.
 
The majority of muon candidates are reconstructed by combining a track
in the ID with a track in the MS. To improve the muon reconstruction
efficiency in the region of $|\eta|<0.1$, which has limited coverage in
the MS, additional muon candidates are identified by matching a
reconstructed ID track to either an MS track segment or a calorimetric
energy deposit consistent with a minimum-ionising particle. In the
region $2.5 < |\eta|< 2.7$, which is not covered by the ID, additional
muons are reconstructed from an MS track with hits in the three MS
layers and combined with forward ID hits, if possible.
Muon candidates are required to satisfy the `loose' criteria defined in
Ref.~\cite{MUON-2018-03} and have $\pT>6\GeV$ and $|\eta|<2.7$. Muons
with an associated ID track must be matched to the primary vertex by having a
longitudinal impact parameter $z_0$ that satisfies $|z_0 \sin(\theta)|<0.5$~mm,
where $\theta$ is the polar angle of the track.
The significance of the transverse impact parameter $d_0$
calculated relative to the measured beam-line position is required
to be $|d_0|/\sigma(d_0) < 3$, where $\sigma(d_0)$ is the uncertainty
in $d_0$.  Furthermore, isolation criteria are applied to suppress non-prompt
muons originating from hadron decays. The isolation selection uses information about ID tracks
and calorimeter energy deposits in a range $\Delta R < 0.2$ around the muon as described
in Ref.~\cite{HIGG-2018-28}.
 
Since muons may lose a significant fraction of their energy by QED FSR,
up to one final-state photon candidate per event is included in the \mmm
calculation to improve the signal reconstruction. Photon candidates
are reconstructed with a procedure similar to the one described in
Ref.~\cite{HIGG-2018-28}, optimised to achieve the best sensitivity
for the \hmm signal. Only photon candidates close to muons
($\Delta R(\gamma, \mu)<0.2$) are considered.
To reduce background from
pile-up interactions, a variable threshold is imposed on the photon
transverse momentum $\pt^\gamma$; the threshold increases
linearly from $\pt^\gamma = 3\GeV$ for $\Delta R = 0$ to
$\pt^\gamma = 8\GeV$ for $\Delta R = 0.2$.
If more than one photon passes this requirement, the photon
with the highest transverse momentum is selected.
A QED FSR candidate is
found in about 5\% of the events and the signal \mmm width is reduced
by about $3\%$ when considering all reconstructed signal events.
With these selections, contributions from the loop-induced decay
$H \to Z\gamma$, $Z\to\mu\mu$~\cite{HIGG-2018-42} are expected to be
around $0.1\%$ of the \hmm yield and are thus neglected in the further
analysis.
 
Electrons are reconstructed by matching clusters of energy in the
electromagnetic calorimeter to tracks in the ID. They are required to
satisfy `Medium' identification criteria~\cite{EGAM-2018-01}, have
$\pT>7\GeV$ and $|\eta| <2.47$ and be outside the region of
$1.37 < |\eta| < 1.52$. Similarly to muons, electrons are required to
be isolated from additional activity measured by ID tracks and
the calorimeters within $\Delta R < 0.2$~\cite{HIGG-2018-28}
and to be matched to the primary vertex
with $|z_0 \sin(\theta)|<0.5$~mm and $|d_0|/\sigma(d_0) < 5$.
 
Jets are reconstructed from `particle flow' objects~\cite{PERF-2015-09}
using the anti-$k_t$ algorithm~\cite{Cacciari:2008gp,Fastjet}
with a radius parameter of $R=0.4$.
Candidate jets must have $|\eta|<4.5$, and the jet \pT must
be larger than $25\,(30)\GeV$ for $|\eta|< 2.4$ ($2.4 <|\eta|<
4.5$). To suppress pile-up contributions, jets with $|\eta|<2.4$ and
$\pT < 60\GeV$ that do not originate from the primary vertex are
rejected using the jet vertex tagging algorithm
(JVT)~\cite{PERF-2014-03}, which combines tracking information into a
multivariate likelihood.
 
Jets containing $b$-hadrons with $|\eta|< 2.5$ are identified as
$b$-tagged jets using a multivariate $b$-tagging algorithm. Two
identification working points (WP) are used~\cite{PERF-2016-05,
FTAG-2018-01} to provide a $60\%$ ($85\%$) efficiency in
$t\bar{t}$ events and a rejection factor of 1200 (25) for
light-flavour jets, respectively.
 
Neutrinos escape from the detector and lead to missing transverse
momentum \etmiss. The \etmiss\ is defined as the magnitude of the
negative vectorial sum of the transverse momenta of the selected and
calibrated physics objects (including muons, electrons and jets) and the ID
tracks not associated with any physics object (soft
term)~\cite{PERF-2016-07, ATLAS-CONF-2018-023}.
 
Events are selected if they contain at least two opposite-charge muon
candidates. The leading muon is required to have $\pT>27\GeV$ to be
above the trigger threshold and in most categories the subleading muon
has to have $\pT>15\GeV$.  Further requirements on the presence or
absence of additional muons, electrons and $b$-tagged jets depend on
the targeted Higgs boson production mode (\ttH, $VH$, or \ggF+VBF),
as detailed in \Sec~\ref{sec:categorisation}.
The main selection
requirements are summarised in \Tab~\ref{tab:evtsel}.
The final signal+background fits are
performed in the region $\mmm = 110$--$160\GeV$, where about $450\,000$
data events are selected. Within the mass window $\mmm = 120$--$130\GeV$,
which contains about $85\%$ of the signal, about 868 \hmm
events are expected. This corresponds to a
total efficiency times acceptance of about $52\%$ with respect to all
\hmm produced in the \ggF, VBF, $VH$ and \ttH processes.
 
\begin{table}[tb]
\begin{center}
\caption{Summary of the main event selection criteria common to
all events as well as the criteria applied to the selection of
hadronic jets.
The bottom
sections give the basic requirements on leptons and $b$-tagged
jets for the analysis categories targeting different Higgs
boson production processes. The subleading muon momentum
threshold is $15\,\GeV$ in all categories except the $VH$
3-lepton categories, where it is lowered to $10\,\GeV$.}
\label{tab:evtsel}
\small
\begin{tabular}{lc}
\toprule
& Selection \\
\midrule
& Primary vertex\\
Common preselection & Two opposite-charge muons\\
& Muons: $|\eta| < 2.7$, $\pT^\mathrm{lead} > 27\GeV$, $\pT^\mathrm{sublead} > 15\GeV$ (except $VH$ 3-lepton)\\
\midrule
Fit Region & $110< \mmm< 160\GeV$\\
\midrule
\multirow{2}{*}{Jets} & $\pT > 25\GeV$ and $|\eta|<2.4$\\
& or with $\pT > 30\GeV$ and $2.4 < |\eta|< 4.5$ \\
\midrule
\ttH Category & at least one additional $e$ or $\mu$ with $\pT > 15\GeV$, at least one $b$-jet ($85\%$ WP)\\
$VH$ 3-lepton Categories & $\pT^\mathrm{sublead} > 10\GeV$, one additional $e$ ($\mu$) with $\pT > 15(10)\GeV$, no $b$-jets ($85\%$ WP)\\
$VH$ 4-lepton Category & at least two additional $e$ or $\mu$ with $\pT > 8, 6\GeV$, no $b$-jets ($85\%$ WP)\\
\ggF+VBF Categories & no additional $\mu$, no $b$-jets ($60\%$ WP)\\
\bottomrule
\end{tabular}
\end{center}
\end{table}
 
 
\section{Event categorisation}
\label{sec:categorisation}
 
Events satisfying the preselection criteria of \Sec~\ref{sec:event} are classified into 20 mutually exclusive categories.
They are defined to exploit the topological and kinematic differences between the background processes and the different
Higgs boson production modes: \ggF, VBF, $VH$ and \ttH.
The background is dominated inclusively by the DY process,
while diboson production, $t \bar{t}$ and single-top production and rarer SM processes such as \ttV
play a significant role in the categories targeting $VH$ and \ttH production.
 
After preselecting events according to the presence of additional leptons and the number of jets and $b$-tagged jets,
boosted decision trees (BDT)~\cite{BDT,Roe:2004na} are trained using the XGBoost package~\cite{xgboost}
to enhance the signal sensitivity as explained in the following.
In order to avoid any potential bias, all trainings are performed using $k$-fold
cross-validation, where $k$ different partitions are used in turn for training, for validation and for testing.
 
The category selections targeting the different Higgs boson production modes are made
in a specific exclusive order, corresponding to the order in which they are presented in
this section.

\subsection{\ttH category}
 
A category enriched in \ttH events is defined in order to target the dileptonic or semileptonic decay of the $t \bar{t}$ system.
Events are considered for this category if there is at least one lepton ($e$ or $\mu$)
with a transverse momentum $\pT > 15\,\GeV$ in addition to the opposite-sign muon pair
and at least one $b$-tagged jet
selected by the 85\%\ efficiency working point.
The two highest-\pT muons with opposite charge are chosen as the Higgs boson decay candidate and used to calculate the variable \mmm used in the final fit. This procedure correctly selects the muon pair coming from the Higgs boson decay in about 80\% of the cases.
After this selection, a BDT is trained using simulated \ttH, \hmm events as signal
and simulated events from all SM background processes as background, both selected in a range $\mmm=100$--$200\,\GeV$.
The 12 variables used in the BDT include the Higgs boson candidate's transverse momentum \ptmm,
the value of the cosine of the lepton decay angle $\cos\theta^*$ in the Collins--Soper frame~\cite{Collins:1977iv},
the transverse momenta of the additional leptons,
the multiplicity of central jets with $|\eta|<2.5$, the multiplicity of $b$-tagged jets, and
the scalar sum of the transverse momenta of all the jets, \HT.
In addition, several invariant masses derived from the reconstructed objects as described in the following
are used.
If there are not enough reconstructed objects in the event to define the invariant masses described in the following,
fixed arbitrary values outside their physical ranges are assigned.
The leptonic top-quark candidate mass \mltop is calculated as the transverse mass of the system composed of the third lepton,
the missing transverse momentum and the $b$-tagged jet candidate (if more than one $b$-tagged jet is present, the one yielding \mltop closest
to $173\,\GeV$ is chosen).
The transverse mass of the leptonic $W$-boson candidate \mlw is calculated from the system composed of the third lepton and the missing transverse momentum.
The hadronic top-quark candidate mass \mhtop is reconstructed from three jets, where one jet must be $b$-tagged (if only one $b$-tagged jet is present, it is used in the reconstruction of both \mltop and \mhtop).
If more than three jets are available, the combination is chosen that maximises the probability of compatibility with
a hadronic top-quark decay in terms of closeness of \mhtop and \mhw to the top-quark and $W$-boson masses, respectively, where the mass of the hadronically decaying $W$ boson, \mhw, is calculated
from the two non-$b$-tagged jets associated with the hadronic top decay.
If, in addition to the two muons associated with the Higgs boson candidate,
two additional opposite-sign and same-flavour leptons are present in the event, their invariant mass is used in the BDT. Similarly, if there are at least three muons reconstructed, the mass of the dimuon pair formed
from the muon with the third-highest \pt and the oppositely charged muon assigned to the Higgs boson is included in the classification.
 
A selection is applied to the BDT score to define one \ttH-enriched category (named \ttH in the following)
by optimising the sensitivity to the predicted SM signal.
The background is expected to be dominated by the \ttZ process with additional contributions from the production
of $t\bar{t}$, dibosons, and \ttH with Higgs boson decays into a final state different from \hmm.
Assuming SM Higgs boson production and decay, 1.2 signal events are expected in this category with a
purity of 98\% for the \ttH process relative to other Higgs boson production modes and
a signal-to-background ratio of 8\% in the mass window $\mmm = 120$--$130\,\GeV$.

\subsection{$VH$ categories}
 
Events not selected in the \ttH category are considered for the
$VH$-enriched categories. The $VH$ categories target signal events
where the Higgs boson is produced in association with a leptonically
decaying vector boson, \wln or \zll with
$\ell=e,\mu$, drastically reducing the DY background.  Events are
required to have no jets identified as $b$-jet candidates by the $b$-tagging algorithm
at the $85\%$ efficiency working point. In addition to a pair of oppositely charged muons from
the \hmm decay, at least one additional isolated muon or electron
must be reconstructed. In events with exactly three leptons, the subleading muon associated
with the Higgs boson candidate is required to have a transverse momentum of at least
10\,\GeV, the additional muon (electron) is required to have a transverse momentum of at least
10 (15)\,\GeV\ and no \zmm candidate, defined as an opposite-charge
dimuon pair with a mass $\mmm=80$--$105\,\GeV$, can be present.
In events with at least four leptons, the two additional leptons (muons or electrons) are required
to have transverse momenta of at least 8 and 6 \GeV\ and at most one \zmm candidate can be present.
If more than two muons are reconstructed, their assignment
to the \hmm and \zmm (\wmn) decay candidates is based on their
charges and on the minimisation of a
$\chi^2$ criterion that takes into account the
difference between the reconstructed and expected (transverse) masses of the two bosons and their expected experimental resolutions.
In the three-lepton case, the $\chi^2$ makes use of the transverse mass
built from the lepton associated with the $W$ boson, called the `$W$ lepton' henceforth, and the missing transverse momentum.
For these topologies the correct pairing is obtained in 93\% and 97\% of the cases for the
four-lepton and three-lepton channels, respectively.
If the additional leptons are electrons, these are
matched without ambiguity to the $W\to e\nu$ or $Z\to ee$ decays.
The two muons chosen as the \hmm candidate are used to calculate the variable \mmm used in the final fit.
 
Two BDTs are trained, separately for the three-lepton and four-lepton events,
to discriminate between the simulated signal and background events satisfying the preselection criteria
and having a dimuon invariant mass in the range $\mmm=110$--$160\,\GeV$.
The three-lepton BDT uses the $WH$, \hmm production as signal.
The variables used include the azimuthal separation $\Delta\phi$ between the Higgs boson candidate and
the missing transverse momentum, the transverse momentum of the $W$ lepton,
the transverse mass of the $W$ boson candidate,
the azimuthal separation and the separation in pseudorapidity $\Delta\eta$ between the
Higgs boson candidate and the $W$ lepton,
the missing transverse momentum, the transverse momentum of the leading jet (if present)
and the number of reconstructed jets.

For the four-lepton events the $ZH$, \hmm production is chosen as signal in the BDT training.
The variables used in the BDT include
the azimuthal separation between the leptons from the \zll candidate,
the azimuthal separation and the separation in pseudorapidity between the
\hmm candidate and the \zll candidate, the invariant mass of the \zll candidate,
the number of jets and the transverse momentum of the two leading jets (if present).
 
Three $VH$ categories are defined by applying selection criteria to
the two BDT scores which optimise the sensitivity to the predicted SM
signal.
Two categories are defined for the three-lepton
events and are named VH3LH and VH3LM, where the former has the higher
signal-to-background ratio.  For the four-lepton events, only one category is
selected and is named VH4L.  The diboson processes are expected to
constitute about 70\% (55\%) of the total background in the VH3LH
(VH3LM) category with smaller contributions expected from top-quark
pair production and the DY process. In the VH4L category, about 98\% of the
background is from the $ZZ$ process. Assuming the SM
Higgs boson production and decay, the numbers of signal events expected
in the $\mmm = 120$--$130$\,\GeV\ mass window for the VH3LM, VH3LH and VH4L
categories are 2.8, 1.4 and 0.5, respectively, and the corresponding
signal-to-background ratios are 0.8\%, 3.7\% and 2.6\%.
The expected \hmm signal purity for the $VH$ production process relative to other Higgs production modes
is 89\% in the VH3LM category and more than 99\% in the VH3LH and VH4L categories.

\subsection{\ggF and VBF categories}
 
The events not selected in the \ttH or $VH$ categories described above,
are further classified according to the number of reconstructed jets into three jet multiplicity
categories: 0-jet, 1-jet and 2-jet, where the last includes events with two or more jets.
Events with at least one $b$-tagged jet selected by the $60\%$ efficiency working point or with a third
muon with $\pt > 15\,\GeV$ are rejected in the \ggF and VBF categories,
as they are found to have a very low signal-to-background ratio and negligible signal sensitivity.
To fully exploit the kinematic differences between the signal and the backgrounds,
which are dominated by DY dimuon production contributing more than 90\% of all background
events after preselection, BDTs are trained in each jet multiplicity category.
All BDTs are trained using the MC background samples and the simulated \hmm signal
in the mass window $\mmm=120$--$130\GeV$.
 
In the 2-jet category, a BDT is trained to disentangle signal
events produced by VBF, used as signal sample in the training, from background events.  This BDT, with a score
denoted by \ovbf, is based on 17 variables related to
the dimuon and dijet systems as described below. The dimuon system is characterised by the
transverse momentum \ptmm, rapidity \ymm and the value of $\cos\theta^*$.
Compared to events from the dominant DY background, signal events
from both \ggF\ and VBF production are characterised by larger \ptmm
and smaller absolute values of \ymm. The $\cos\theta^*$ distributions
provide some discrimination due to different spin-structures and $Z$--$\gamma$ interference effects
for the DY background~\cite{STDM-2016-04}.  For the leading and subleading jets in the event
(denoted by $j_1$ and $j_2$), the following variables are computed:
\pt and $\eta$ of $j_1$ and $j_2$; the azimuthal separation between the dimuon system and each jet, $\Delta\phi_{\mu\mu,j_1}$ and
$\Delta\phi_{\mu\mu,j_2}$; the kinematics of the dijet system ($jj$)
characterised by transverse momentum $p_\mathrm{T}^{jj}$, mass $m_{jj}$ and rapidity $y_{jj}$;
and the azimuthal separation between the dimuon and dijet  systems,
$\Delta\phi_{\mu\mu,jj}$. These variables exploit the unique signature of the
VBF process: two high-\pt jets separated by a large rapidity gap with little hadronic activity.
For jets that have a high $\pT > 50\,\GeV$ and are in the central region $|\eta| < 2.1$, the multiplicity of
ID tracks with $\pT > 0.5\,\GeV$ associated with each of the two leading jets,
$N^{j_i}_\mathrm{track}$, is also used to help
discriminate between jets produced by fragmentation of gluons and quarks~\cite{ATL-PHYS-PUB-2017-009, STDM-2015-12}.
In addition,
\etmiss and
$H_\mathrm{T}$, which can discriminate the
$t\bar{t}$ background from the signal, are also considered.
 
From the \ovbf classifier, four categories in the region with the highest
score are selected and named VBF Very High, High, Medium, and Low.
The expected \hmm signal contributing to these categories is
dominated by the VBF process with a purity ranging from about 93\%
(VBF Very High) to 65\% (VBF Low) and the expected SM
signal-to-background ratio computed in the $120$--$130$\,\GeV\ mass
window varies between 18\% (VBF Very High) and 2.8\% (VBF Low).
The predicted number of SM signal events in the VBF categories ranges
between 2.8 and 7.5 events.
 
The remaining events are  considered for
further classification in three other BDTs split by jet multiplicity, with scores denoted as \oggfi .
These BDTs are trained with both the \hmm\ \ggF
and VBF production MC samples as signal. In each jet category, the same three
variables characterising the dimuon system and discussed above are
used, i.e.\ \ptmm, \ymm and $\cos\theta^*$.  For the 1-jet BDT, they
are complemented by the transverse momentum $\pt^{j_1}$,
pseudorapidity $\eta_{j_1}$, the track multiplicity $N^{j_1}_\mathrm{track}$ in the jet,
and the azimuthal separation from the
dimuon system, $\Delta\phi_{\mu\mu,j_1}$. In the
2-jet category the same variables as in the VBF BDT are
used.
 
The events in each of these three jet multiplicity categories are further classified
into four categories on the basis of the \oggfi\ scores to
yield a total of 12 mutually exclusive categories called $N$-jet
Very High, High, Medium and Low with $N=0,1,2$. The \ggF production process
contributes $80$\%--$100\%$ of the expected \hmm signal in these
categories. The expected SM signal-to-background ratio
in the $120$--$130$\,\GeV\ mass window varies from $1.7$\%--$1.5$\%
(2-jet Very High and 1-jet Very High) to $0.07$\% (0-jet Low).
The predicted number of SM signal events in the \ggF\ categories ranges
between 17 and 125 events.
 
\section{Signal and background modelling and systematic uncertainties}
\label{sec:modelling}
 
The signal extraction is based on a binned maximum-likelihood fit to the invariant mass spectrum of the dimuon system
as described in \Sec~\ref{sec:result}.
Analytic models are used in the fit to describe the \mmm\ distributions for both the signal
and background processes.
 
\subsection{Signal modelling}
 
In the SM, the $H\to \mu\mu$ signal is predicted to be a narrow resonance with a width of $4.1\MeV$ for $m_{H}=125.09\GeV$.
The observed signal shape is thus determined by detector resolution effects on the muon momentum measurement.
A double-sided Crystal Ball function is used for the Higgs signal model.
This function is a modification of the Crystal Ball function~\cite{Oreglia:1980cs, HIGG-2014-04}, and consists of a Gaussian central part
with a power-law tail on each side.
 
For each of the 20 categories, the signal parameters are fitted to signal MC spectra summed over all production modes (\ggF, VBF, $VH$, \ttH)
assuming the relative normalisations as predicted by the SM.
Within each category no significant differences are found between the signal shapes of the different production modes,
indicating that the signal parameterisation is not sensitive to the assumption on their relative normalisations.
The width of the Gaussian component of the double-sided
Crystal Ball function varies between 2.6 and 3.2\,\GeV\ depending on the category.
Potential biases in the extracted signal
yields due to the analytic parameterisations are tested with a
signal injection procedure: in a signal-plus-background fit to
pseudo-data constructed from the expected signal and background distributions,
the extracted signal yields agree with those
injected within the statistical accuracy of about 0.3\%.
 
Several sources of systematic uncertainty in the signal modelling
are considered, including both theoretical and experimental
effects.
The theoretical uncertainties in the signal production affect the
number of signal events expected in each category. The
uncertainties considered for the main production modes (\ggF and VBF)
include the impact of the missing higher-order QCD corrections, PDFs, the underlying event and
hadronisation.  In particular, the uncertainty in the \ggF signal is derived using the
approach described in Ref.~\cite{deFlorian:2016spz}, including effects from the variation of
QCD scales for factorisation, renormalisation and resummation,
and the migration between jet-multiplicity regions~\cite{Liu:2013hba, Boughezal:2013oha, Campbell:2010ff, Stewart:2011cf, Gangal:2013nxa, Frederix:2016cnl, deFlorian:2012mx, Grazzini:2013mca}.
The uncertainty in the \ggF\ Higgs boson transverse momentum,
including the effect of migration between different kinematic regions and
of the treatment of the top-quark mass in the loop corrections,
is also taken into account. In addition,
dedicated uncertainties are assigned for the \ggF\ signal acceptance in VBF topologies.
The uncertainties in the predicted SM branching ratio and
Higgs boson production cross sections are included in accord with
Ref.~\cite{deFlorian:2016spz}.
The uncertainties associated with the modelling
of the underlying event and parton showering are estimated by considering the \PYTHIAV{8} systematic
eigentune variations and by comparing events showered by \PYTHIAV{8} with those showered by
\HERWIGV{7}~\cite{Bahr:2008pv,Bellm:2015jjp}. The impact of the theory
uncertainties on the predicted signal acceptances in the different
categories ranges between a few per mill and 15\% for
\ggF\ production. Similarly, for the VBF production the impact of the
theory uncertainties on the predicted signal acceptances varies between a few per
mill and 7\%. For the $VH$ and \ttH categories
the theory systematic uncertainties have an impact on the predicted signal acceptances
of between a few per mill and about 18\%.
 
Systematic uncertainties related to the different reconstructed
physics objects used in the analysis affect the expected signal yields
in each category. In addition, systematic uncertainties in the
muon momentum scale and resolution also affect the signal mass
distribution. The experimental uncertainties considered are the muon
reconstruction and identification efficiencies, the efficiencies due
to the trigger, isolation and impact parameter requirements, the muon
momentum scale and resolution~\cite{MUON-2018-03,PERF-2015-10,TRIG-2018-01}, the determination of the \etmiss\ soft
term~\cite{PERF-2016-07}, the $b$-tagging efficiency~\cite{FTAG-2018-01}, the uncertainty in the number of tracks
associated with the jets~\cite{ATL-PHYS-PUB-2017-009}, the pile-up modelling~\cite{ATL-PHYS-PUB-2016-017}, and uncertainties in the electron
reconstruction and identification efficiency~\cite{EGAM-2018-01}
as well as in the jet reconstruction efficiency, energy scale and resolution~\cite{PERF-2016-04}.
The impact of the experimental uncertainties on the predicted signal yields and modelling
in the different categories is dominated by the uncertainties in the jet energy scale
and resolution and the muon momentum resolution. The former can affect
signal yields by up to about 10\% in some of the 2-jet categories. The
muon momentum resolution uncertainty has an impact on the fitted
yields ranging between 1\% and 6\% depending on the category.
 
The experimental uncertainty of $240 \MeV$ in the assumed value of the Higgs
mass from Ref.~\cite{HIGG-2014-14} is also taken into account.
All these sources of uncertainty are included in the signal
extraction fit described in \Sec~\ref{sec:result} through nuisance
parameters acting on the relative signal yields in the different
categories and on the signal mass distributions.

\subsection{Background modelling}
 
Due to the very small signal-to-background ratio, which is
at the level of $0.2\%$ in the region $\mmm = 120$--$130\GeV$ in an inclusive selection,
an accurate determination of the background is of paramount importance.
The $\mmm$ background spectrum is parameterised by analytic
functions that can describe this distribution at the per-mill level to
avoid a significant bias in the extracted signal yields.
The mass range used for the fit, $\mmm = 110$--$160\GeV$, is optimised
to obtain the best signal sensitivity taking into account the statistical and systematic uncertainties.

For the \ggF and VBF categories, the background is dominated by the DY process,
which accounts for more than 90\%\ of the total, with small contributions from
top-quark processes (mainly in the 2-jet categories) and diboson
production. In the \ttH and $VH$ categories,
the dominant backgrounds are associated production of \ttZ and $VZ$ with \zmm, respectively, while the DY process, $t\bar{t}$ production and other diboson processes give minor contributions.
 
To achieve the required accuracy in the analytic description
of the background  $\mmm$ distribution, the following approach is used. A
\textit{core} function that describes the DY mass shape inclusively is
multiplied by an \textit{empirical} function that can correct
for distortions of the mass shape due to the event selection and
categorisation, higher-order theory corrections and other smaller background contributions. The empirical functions chosen are also flexible enough to describe the background shape in categories where the dominant background is not the DY process.
The core function has no free parameters and is common to all categories, while
the empirical functions have a certain number of free parameters that
are selected and fit to data independently in each category.
 
The core component of the background is an analytical function based on a LO DY line-shape,
described in Appendix~\ref{app:dyls},
convoluted with detector effects.
The experimental resolution in the dimuon invariant mass is found to
have an important effect on the core function, since it produces a
significant shape variation in the mass region just above the
$Z$-boson resonance and thus influences the lower end of the fit
region in the \hmm search. To take this effect into account, the LO DY
line-shape is convolved with a Gaussian function with a mass-dependent
resolution derived from the simulation.
 
The core function is multiplied by the empirical component to obtain
the final background parameterisation used in the fits to the \mmm
spectra. Two families of functions are studied for this empirical
component: power-law functions (`Power') and exponentials of
polynomials (`Epoly'), as defined in \Tab{~\ref{tab:bkgmodelling:pdflist}}.
 
\begin{table}[htb]
\centering
\caption{List of tested empirical functional forms for the background modelling.}
\label{tab:bkgmodelling:pdflist}
\begin{tabular}{l c}
\toprule
{{Function}} & {{Expression}} \\
\midrule
{PowerN}      &    $  \mmm^{(a_{0}  + a_{1} \mmm+  a_{2} \mmm^2 + ... +a_{N} \mmm^N) }  $ \\
{EpolyN}      &    $ \exp(a_{1} \mmm + a_{2} \mmm^2 + ... + a_{N}  \mmm^N )  $ \\
\bottomrule
\end{tabular}
\end{table}
 
The criteria used to select the background functions from among those
listed above and to determine the associated systematic uncertainty,
referred to as the spurious signal (SS)~\cite{HIGG-2012-27}, are described in the following.
The SS yields are taken as the measured signals obtained in signal-plus-background fits
to the background-only MC templates. They are determined not only for a signal
mass of $125\,\GeV$, but also for values of $m_H$ between 120 and 130\,\GeV\ in
steps of $1\,\GeV$. The templates derived from fast and full simulation DY samples are reweighted using first- or second-order polynomial functions
in \mmm to the data sidebands for all these studies.
 
As a first requirement, only functions able to fit the data sidebands,
the fully simulated background samples and the fast DY simulation
with a $\chi^{2}$ probability of the fit greater than $1\%$ (for all these samples) are considered.
For the \ttH and $VH$ categories, only the data sidebands
and the fully simulated background samples are considered for these criteria,
and the DY contribution is neglected since it is very small and subject to large statistical fluctuations.
 
For the functions that satisfy these criteria, a spurious-signal test
is performed separately in each category.  For the \ggF and VBF-enriched
categories the primary fast-simulation DY sample based on \SHERPA as
described in Section~\ref{sec:samples} is used, since it has high
statistical precision, while for the \ttH and the $VH$
categories the fully simulated non-DY background samples are used.
Only the functions with the absolute value of the SS
below 20\% of the expected signal statistical error in data in the
mass range 120 to 130\,\GeV\ are considered.
When applying this requirement, the MC statistical error is subtracted from the absolute value of the SS.
Among the functions that pass this requirement, those with the smallest number of
degrees of freedom are selected in each category to minimise the statistical uncertainty that dominates in this
search.  If more than one function per category passes this last selection, the one with
the smallest SS is selected. The maximum absolute value of the SS in the mass
range 120--130\,\GeV\ is taken to be the background modelling
uncertainty for the respective category.
 
As an additional cross-check, the SS tests for the \ggF and
VBF-enriched categories are also performed on the fully simulated \SHERPA DY samples
and the alternative fast DY
simulation based on the merged \POWHEGBOX and \ALPGEN DY samples as
explained in Section~\ref{sec:samples}. Further cross-checks are
performed with the fast DY simulation after applying several
theoretical variations, such as changes of the QCD renormalisation and
factorisation scales by factors of two and one half and alternative
PDF sets, and experimental variations of the muon momentum resolution
and scale and the pile-up jet modelling within the experimental
uncertainties. In all these checks, no statistically significant increase in the SS
values is found, hence they are not included as additional systematic uncertainty
since their impact would be negligible.
The SS systematic uncertainty also addresses any potential local biases
in the mass spectra close to the signal region caused by the experimental selections,
such as the BDT score requirements or the lepton pairing procedure in the $VH$ categories.
 
After applying the above criteria, there is no evidence of
statistically significant mismodelling, as no SS values are found that
are more than two standard deviations away from zero for a signal mass
of $125\GeV$. This considers the statistical accuracy of the fast DY
simulation that is about ten times better than that of the data. All
the SS are considered as uncorrelated systematic uncertainties among
the different categories. If the SS uncertainties were
considered as fully correlated between categories, the expected
significance would change by less than 2\%.
 
The SS uncertainties in the different categories range from a few per cent up to about 20\%
of the expected data statistical uncertainties in the VBF and \ggF categories and
up to about 30\% in the $VH$ and \ttH categories, which have less statistical
precision in their background simulated samples.

\subsection{Other systematic uncertainties}
 
In addition to the systematic uncertainties in the signal and
background modelling described above, the uncertainty of $1.7\%$ in the
combined 2015--2018 integrated luminosity is also considered. It
is derived from the calibration of the luminosity scale using $x$--$y$
beam-separation scans~\cite{ATLAS-CONF-2019-021}, obtained using the LUCID-2 detector \cite{LUCID2} for the primary luminosity measurements.
 
\section{Results}
\label{sec:result}
 
The signal yield is obtained by a simultaneous binned maximum-likelihood fit to the \mmm\ distributions of the 20
categories in the range 110--160\,\GeV . The chosen bin size is 0.1\,\GeV .
Confidence intervals are based on the profile-likelihood-ratio test statistic~\cite{Cowan:2010st}.
The systematic uncertainties listed in \Sec{~\ref{sec:modelling}} are implemented in the fit as nuisance
parameters constrained by additional Gaussian or log-normal likelihood terms
and the Higgs boson is assumed to have a mass of $m_H = (125.09 \pm 0.24) \GeV$.
 
The best-fit value of the signal strength parameter,
defined as the ratio of the observed signal  yield to the one
expected in the SM, is
$\mus = 1.2 \pm 0.6$, corresponding to an observed (expected)
significance of $2.0 \sigma$ ($1.7 \sigma$) with respect to the
hypothesis of no \hmm\ signal. The spectra of the dimuon invariant mass for  all the analysis
categories after the signal-plus-background fit
are presented in \Fig~\ref{fig:Inclusive}.
In \Fig~\ref{fig:InclusiveW} the events are weighted by
$\ln(1 + S/B)$, where $S$  are the observed signal yields 
and $B$ are the background yields
derived from the fit to data in the $\mmm = 120$--$130\,\GeV$ window.
These values for $S$, $B$ and other key quantities are listed in \Tab{~\ref{tab:Results}}.
 
The best-fit values of the signal
strength parameters for the five major groups of categories (\ttH+$VH$, \ggF 0-jet, 1-jet, 2-jet, and VBF)
are shown in \Fig~\ref{fig:5categories} together with the combined value.
A goodness-of-fit test is performed using the saturated model technique~\cite{Cousin2013} and
returns a probability of 10\%.
 
\begin{figure}[htbp]
\centering
 
\subfloat[ ]{\includegraphics[width=0.48\textwidth]{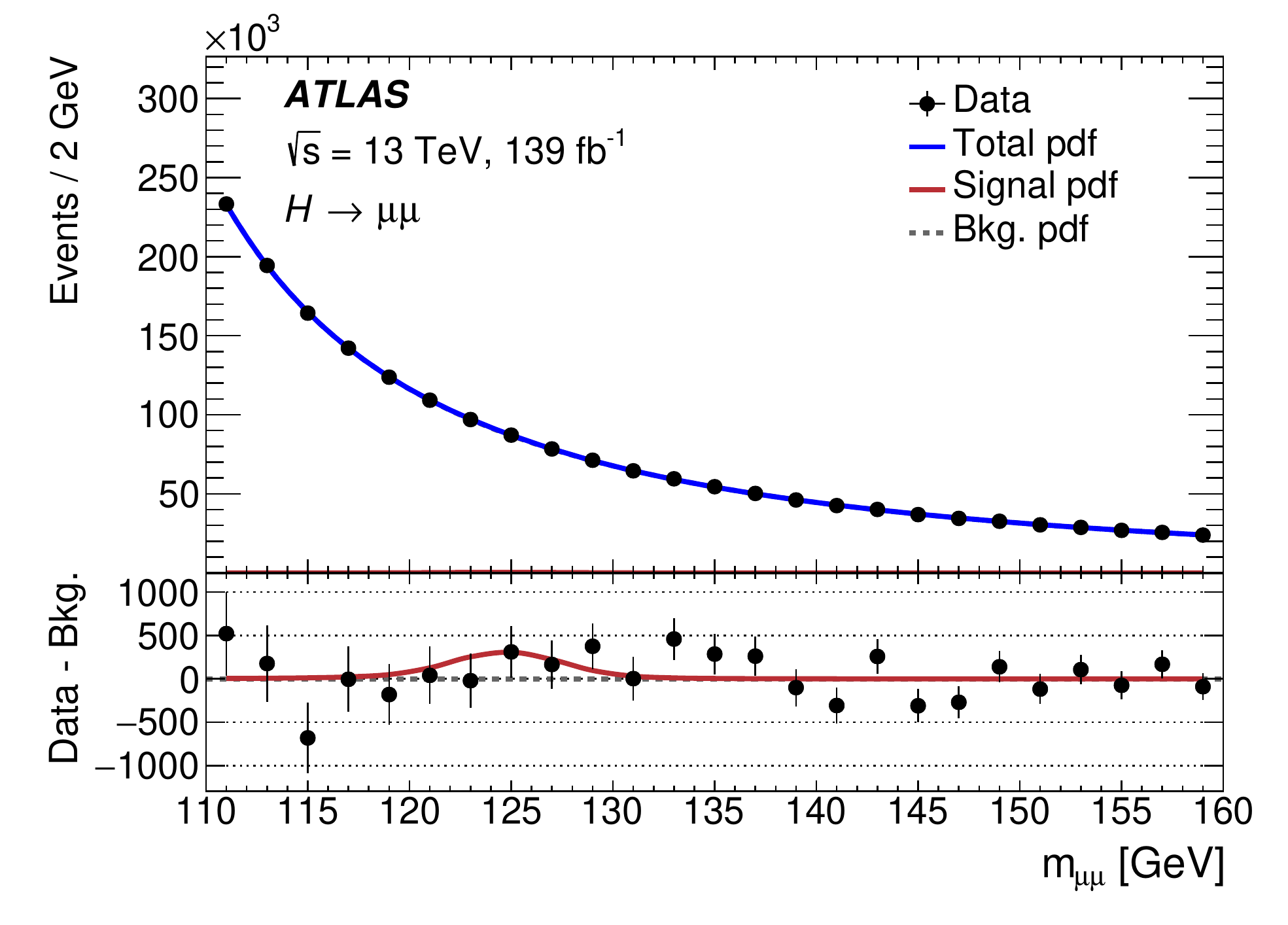}}
\subfloat[ ]{\includegraphics[width=0.48\textwidth]{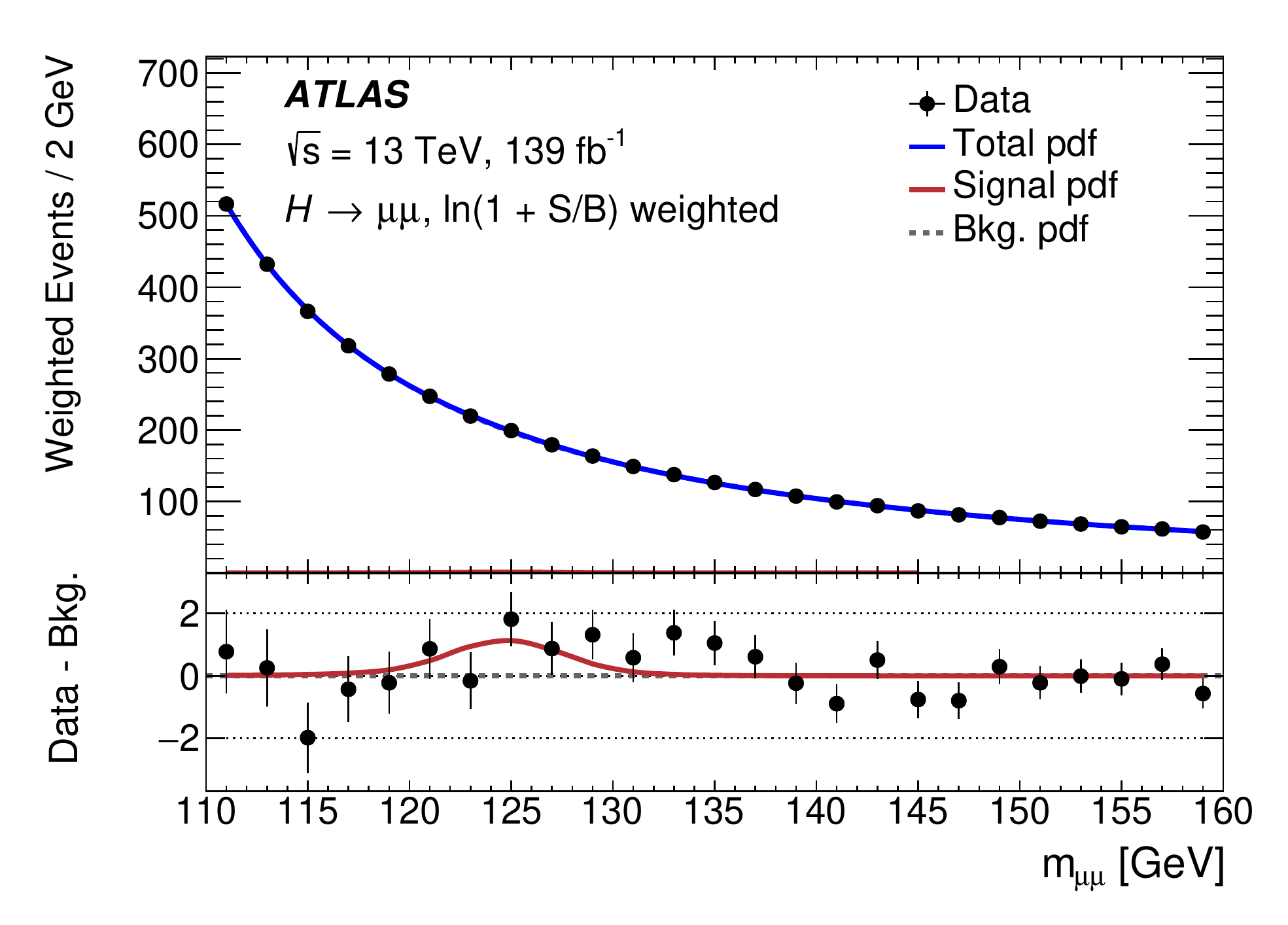}\label{fig:InclusiveW}}
\caption{ Dimuon invariant mass spectrum in all the analysis
categories observed in data. In (a) the unweighted sum of all events
and signal plus background probability density functions (pdf) are shown, while in (b) events and pdfs are
weighted by $\ln(1 + S/B)$, where $S$ are the observed signal yields
and $B$ are the background yields
derived from the fit to data in the $\mmm = 120$--$130\,\GeV$ window.
The background and signal pdf are derived from the fit to the data, with $S$ normalised to its best-fit value.
The lower panels compare the fitted signal pdf, normalised to the signal best-fit value,
to the difference between the data and the background model.
The error bars represent the data statistical uncertainties.
}
\label{fig:Inclusive}
\end{figure}

\begin{table}[htbp]
\centering
\caption{Number of events observed in the $\mmm = 120$--$130\GeV$ window in data, the number of signal events expected in the SM ($S_\mathrm{SM}$), and
events from signal ($S = \mu \times S_\mathrm{SM}$) and background ($B$) as derived from the combined fit to the data with a signal strength parameter of $\mu = 1.2$.
The uncertainties in $S_\mathrm{SM}$ correspond to the systematic uncertainty of the SM prediction,
the uncertainty in $S$ is given by that in $\mu$, and the uncertainty in $B$
is given by the sum in quadrature of the statistical uncertainty from the fit and the SS uncertainty.
In addition the observed number of signal events divided by the square root of the number of background events ($S/\sqrt{B}$)
and the signal-to-background ratio ($S/B$) in \% for each of the 20 categories described in the text are displayed.
In the last column, the width of the Gaussian component of the double-sided Crystal Ball function used
in the signal modelling ($\sigma$, as described in Section~\ref{sec:modelling}) is reported. }
\label{tab:Results}
\small
\begin{tabular}{ l  r  c c c c c c}
\toprule
Category & Data &  $S_\mathrm{SM}$  & $S$        & $B$       & $S/\sqrt{B}$ & $S/B$ [\%] &  $\sigma ~[\gev] $\\
\midrule
VBF Very High     & 15    & 2.81 $\pm$ 0.27 & 3.3 $\pm$ 1.7  & 14.5    $\pm$ 2.1 & 0.86 & 22.6~~  & 3.0\\
VBF High          & 39    & 3.46 $\pm$ 0.36 & 4.0 $\pm$ 2.1  & 32.5    $\pm$ 2.9 & 0.71 & 12.4~~  & 3.0\\
VBF Medium        & 112   & 4.8  $\pm$ 0.5  & 5.6 $\pm$ 2.8  & 85      $\pm$ 4   & 0.61 & 6.6     & 2.9 \\
VBF Low           & 284   & 7.5  $\pm$ 0.9  & 9   $\pm$ 4    & 273     $\pm$ 8~~   & 0.53 & 3.2   & 3.0 \\
2-jet Very High   & 1030  & 17.6 $\pm$ 3.3~~  & 21  $\pm$ 10   & 1024    $\pm$ 22~~  & 0.63 & 2.0 & 3.1\\
2-jet High        & 5433  & 50   $\pm$ 8~~    & 58  $\pm$ 30   & 5440    $\pm$ 50~~  & 0.77 & 1.0 & 2.9\\
2-jet Medium      & 18\,311 & 79   $\pm$ 15   & 90  $\pm$ 50   & 18\,320   $\pm$ 90~~~~~  & 0.66 & 0.5 & 2.9 \\
2-jet Low         & 36\,409 & 63   $\pm$ 17   & 70  $\pm$ 40   & 36\,340   $\pm$ 140~~~ & 0.37 & 0.2   & 2.9\\
1-jet Very High   & 1097  & 16.5 $\pm$ 2.4~~  & 19  $\pm$ 10   & 1071    $\pm$ 22~~  & 0.59 & 1.8      & 2.9 \\
1-jet High        & 6413  & 46   $\pm$ 7~~    & 54  $\pm$ 28   & 6320    $\pm$ 50~~  & 0.69 & 0.9 & 2.8 \\
1-jet Medium      & 24\,576 & 90   $\pm$ 11   & 100 $\pm$ 50~~   & 24\,290   $\pm$ 100~~~ & 0.67 & 0.4 & 2.7 \\
1-jet Low         & 73\,459 & 125  $\pm$ 17~~   & 150 $\pm$ 70~~   & 73\,480   $\pm$ 190~~~ & 0.53 & 0.2 & 2.8 \\
0-jet Very High   & 15\,986 & 59   $\pm$ 11   & 70  $\pm$ 40   & 16\,090   $\pm$ 90~~~~~  & 0.55 & 0.4   & 2.6 \\
0-jet High        & 46\,523 & 99   $\pm$ 13   & 120 $\pm$ 60~~   & 46\,190   $\pm$ 150~~~ & 0.54 & 0.3   & 2.6 \\
0-jet Medium      & 91\,392 & 119  $\pm$ 14~~   & 140 $\pm$ 70~~   & 91\,310   $\pm$ 210~~~ & 0.46 & 0.2 & 2.7 \\
0-jet Low         & 121\,354 & 79   $\pm$ 10   & 90  $\pm$ 50   & 121\,310  $\pm$ 280~~~~~ & 0.26 & 0.1  & 2.7 \\
VH4L              & 34    & 0.53 $\pm$ 0.05 & 0.6 $\pm$ 0.3  & 24      $\pm$ 4   & 0.13 & 2.6 & 2.9 \\
VH3LH             & 41    & 1.45 $\pm$ 0.14 & 1.7 $\pm$ 0.9  & 41      $\pm$ 5   & 0.27 & 4.2 & 3.1 \\
VH3LM             & 358   & 2.76 $\pm$ 0.24 & 3.2 $\pm$ 1.6  & 347     $\pm$ 15  & 0.17 & 0.9 & 3.0 \\
\ttH              & 17    & 1.19 $\pm$ 0.13 & 1.4 $\pm$ 0.7  & 15.1    $\pm$ 2.2 & 0.36 & 9.2 & 3.2 \\
\bottomrule
\end{tabular}
\end{table}

\begin{figure}[htbp]
\centering
\includegraphics[width=0.7\textwidth]{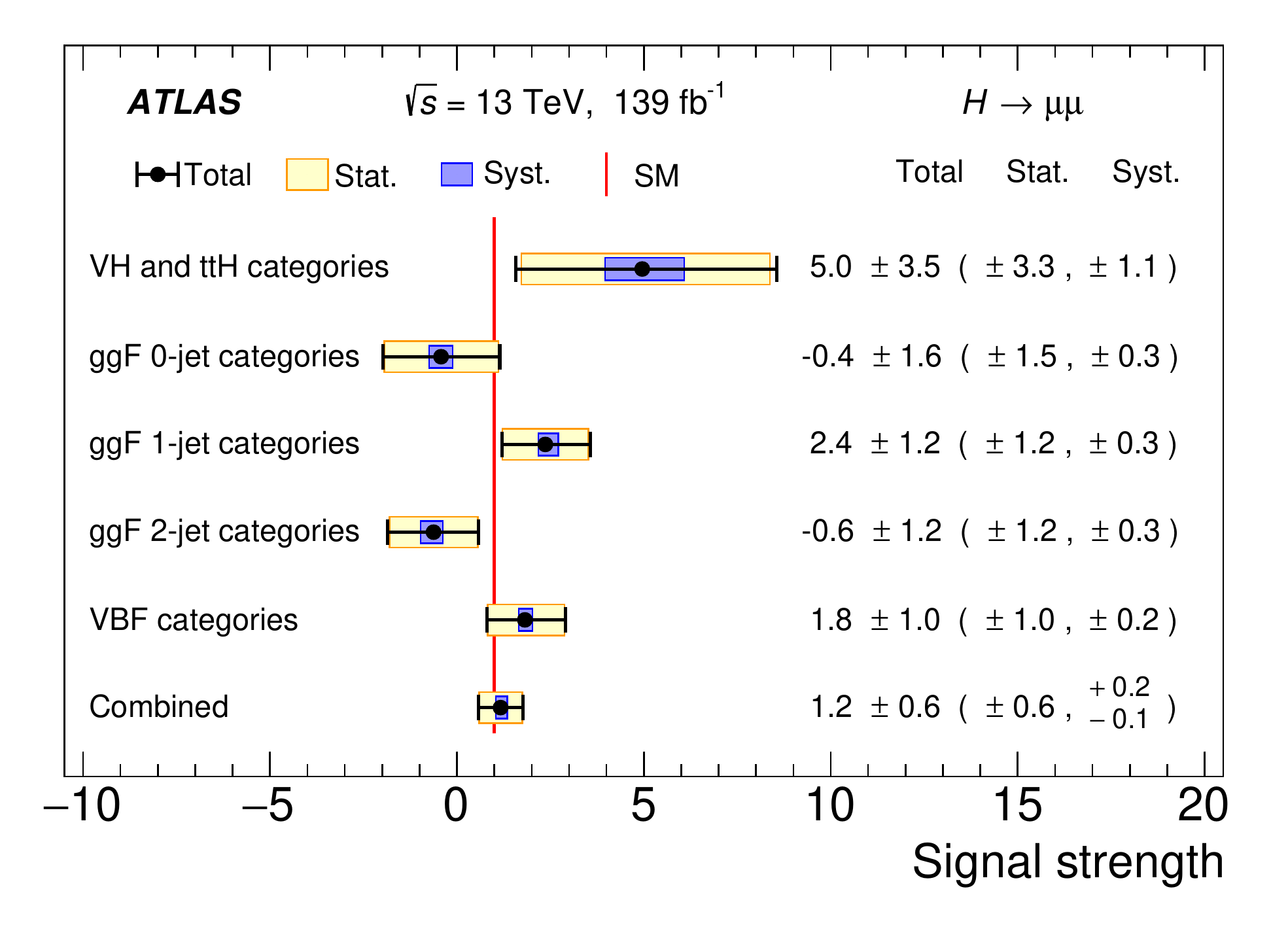}
\caption{ The best-fit values of the signal strength parameters for the five major groups of categories (\ttH+$VH$, \ggF 0-jet, 1-jet, 2-jet, and VBF)
together with the combined value.}
\label{fig:5categories}
\end{figure}

The signal strength uncertainty is dominated by the data statistical error of about $\pm 0.58$.
The impact of the systematic uncertainties on the signal strength is found to be $^{+0.18}_{-0.13}$,
with contributions from the signal theory uncertainties that account for $^{+0.13}_{-0.08}$,
the signal experimental uncertainties that account for $^{+0.07}_{-0.03}$ and the spurious-signal
uncertainties that account for $\pm 0.10$.
 
The compatibility of the measured signal strengths between the 20
categories
is tested by repeating the fit after allowing each category to have
its own signal strength parameter. The probability of compatibility is found to be at the level of 2\%.
With the same methodology, the probability of compatibility between the signal strengths of
the five groups of categories shown in \Fig~\ref{fig:5categories}
is found to be 20\%.
Among the 20 categories, those with an individual signal strength at the level of
two standard deviations from the mean value are the VBF Medium, the 0-jet Very High and the VH4L.
For each of these three categories, it was checked that adding one degree of freedom
to the function used to model the background or changing the functional form from `Power' to `Epoly'
does not significantly impact the analysis results or the probability of compatibility between the 20 categories.
 
An upper limit on the signal strength \mus\ is computed using a modified frequentist CL$_\text{s}$
method~\cite{Cowan:2010st,Read:2002hq}. The observed upper limit on \mus\ at $95\%$ confidence level (CL) is
found to be $2.2$, with an expected limit of $1.1$ for the case of no \hmm\ signal and an
expected limit of $2.0$ for the case of an \hmm\ signal at SM strength.
The corresponding branching ratio upper limit at  $95\%$ CL
is $B(\hmm) < 4.7 \cdot 10^{-4}$, assuming the SM cross section for Higgs boson production.
 
This result represents an improvement of about a factor of 2.5 in expected
sensitivity compared with the previous ATLAS publication~\cite{HIGG-2016-10}.
Of this improvement, a factor of about two is due to the larger analysed dataset and
the additional 25\% improvement can be attributed to more advanced analysis techniques.
 

\section{Conclusion}
 
A search for the rare dimuon decay of the Higgs boson is performed
using the full Run~2 dataset of \lumi\ collected with the ATLAS
detector in $pp$ collisions at $\sqrt{s}=13\TeV$ at the LHC.
The best-fit value for the SM \hmm\ signal strength parameter is found
to be $\mus = 1.2 \pm 0.6$, corresponding to an
observed (expected) significance of $2.0 \sigma$ ($1.7 \sigma$) with respect to the hypothesis
of no \hmm\ signal.
An upper limit of $2.2$ at $95\%$ CL is set on the signal strength, while $1.1$ is expected
for the case of no \hmm\ signal.
 
\section*{Acknowledgements}

We thank CERN for the very successful operation of the LHC, as well as the
support staff from our institutions without whom ATLAS could not be
operated efficiently.
 
We acknowledge the support of ANPCyT, Argentina; YerPhI, Armenia; ARC, Australia; BMWFW and FWF, Austria; ANAS, Azerbaijan; SSTC, Belarus; CNPq and FAPESP, Brazil; NSERC, NRC and CFI, Canada; CERN; ANID, Chile; CAS, MOST and NSFC, China; COLCIENCIAS, Colombia; MSMT CR, MPO CR and VSC CR, Czech Republic; DNRF and DNSRC, Denmark; IN2P3-CNRS and CEA-DRF/IRFU, France; SRNSFG, Georgia; BMBF, HGF and MPG, Germany; GSRT, Greece; RGC and Hong Kong SAR, China; ISF and Benoziyo Center, Israel; INFN, Italy; MEXT and JSPS, Japan; CNRST, Morocco; NWO, Netherlands; RCN, Norway; MNiSW and NCN, Poland; FCT, Portugal; MNE/IFA, Romania; JINR; MES of Russia and NRC KI, Russian Federation; MESTD, Serbia; MSSR, Slovakia; ARRS and MIZ\v{S}, Slovenia; DST/NRF, South Africa; MICINN, Spain; SRC and Wallenberg Foundation, Sweden; SERI, SNSF and Cantons of Bern and Geneva, Switzerland; MOST, Taiwan; TAEK, Turkey; STFC, United Kingdom; DOE and NSF, United States of America. In addition, individual groups and members have received support from BCKDF, CANARIE, Compute Canada, CRC and IVADO, Canada; Beijing Municipal Science \& Technology Commission, China; COST, ERC, ERDF, Horizon 2020 and Marie Sk{\l}odowska-Curie Actions, European Union; Investissements d'Avenir Labex, Investissements d'Avenir Idex and ANR, France; DFG and AvH Foundation, Germany; Herakleitos, Thales and Aristeia programmes co-financed by EU-ESF and the Greek NSRF, Greece; BSF-NSF and GIF, Israel; La Caixa Banking Foundation, CERCA Programme Generalitat de Catalunya and PROMETEO and GenT Programmes Generalitat Valenciana, Spain; G\"{o}ran Gustafssons Stiftelse, Sweden; The Royal Society and Leverhulme Trust, United Kingdom.
 
The crucial computing support from all WLCG partners is acknowledged gratefully, in particular from CERN, the ATLAS Tier-1 facilities at TRIUMF (Canada), NDGF (Denmark, Norway, Sweden), CC-IN2P3 (France), KIT/GridKA (Germany), INFN-CNAF (Italy), NL-T1 (Netherlands), PIC (Spain), ASGC (Taiwan), RAL (UK) and BNL (USA), the Tier-2 facilities worldwide and large non-WLCG resource providers. Major contributors of computing resources are listed in Ref.~\cite{ATL-SOFT-PUB-2020-001}.
 
 
\appendix
\section{Expression for the leading-order Drell--Yan lineshape}
\label{app:dyls}
 
The core component of the background function is based on a LO
DY line-shape (see e.g. Ref.~\cite{STDM-2016-04}):
\begin{equation*}
DY(m_{\mu\mu}) = \sum_q{\mathcal{L}_{q\bar{q}}(m_{\mu\mu})\cdot \sigma_{q\bar{q}}(m_{\mu\mu}) }\;,\; q=u,\; s,\; d\,.
\end{equation*}
 
The parton luminosity contribution $\mathcal{L}_{q\bar{q}}$
is derived from the PDF4LHC15 PDF set
as a function of $\hat{s}=\mmm^{2}$ using
APFEL~\cite{Bertone:2013vaa} interfaced to
LHAPDF~\cite{Buckley:2014ana} and parameterised using a 6th order
polynomial. The matrix element component
$\sigma_{q\bar{q}}(\hat{s}) = \sigma_{q\bar{q}}(m_{\mu\mu}) / (2
m_{\mu\mu})$ can be expressed as
\begin{align*}
\sigma_{q\bar{q}}(\hat{s}) &=\frac{4\pi\alpha^2}{3 \hat{s} N_c}[Q_q^2-2Q_qV_\ell V_q\chi_{Z\gamma}(\hat{s})+(A_\ell^2+V_\ell^2)(A_q^2+V_q^2)\chi_Z(\hat{s})]\,,\\
\end{align*}
where
\begin{align*}
\chi_{Z\gamma}(\hat{s}) &=\kappa\frac{\hat{s}(\hat{s}-m_Z^2)}{(\hat{s}-m_Z^2)^2+\Gamma_Z^2m_Z^2}\,,\\
\chi_Z(\hat{s}) &=\kappa^2\frac{\hat{s}^2}{(\hat{s}-m_Z^2)^2+\Gamma_Z^2m_Z^2}\,,\\
\kappa &=\frac{\sqrt{2}G_Fm_Z^2}{4\pi\alpha}\,.
\end{align*}
Here $Q,V,A$ denote the electric charges, vector and axial-vector
couplings of the fermions, $\alpha, G_F$ the electroweak couplings,
$m_Z, \Gamma_Z$ the mass and width of the $Z$-boson using values from Ref.~\cite{pdg2018} and
$N_c=3$ the number of QCD colour charges.
The DY function described above is then convolved with a Gaussian
function with a mass-dependent resolution derived from the simulation.
 
 
\printbibliography
 
\clearpage

\clearpage
 
\begin{flushleft}
\hypersetup{urlcolor=black}
{\Large The ATLAS Collaboration}

\bigskip

\AtlasOrcid[0000-0002-6665-4934]{G.~Aad}$^\textrm{\scriptsize 102}$,    
\AtlasOrcid[0000-0002-5888-2734]{B.~Abbott}$^\textrm{\scriptsize 128}$,    
\AtlasOrcid{D.C.~Abbott}$^\textrm{\scriptsize 103}$,    
\AtlasOrcid[0000-0002-2788-3822]{A.~Abed~Abud}$^\textrm{\scriptsize 36}$,    
\AtlasOrcid[0000-0002-1002-1652]{K.~Abeling}$^\textrm{\scriptsize 53}$,    
\AtlasOrcid[0000-0002-2987-4006]{D.K.~Abhayasinghe}$^\textrm{\scriptsize 94}$,    
\AtlasOrcid[0000-0002-8496-9294]{S.H.~Abidi}$^\textrm{\scriptsize 167}$,    
\AtlasOrcid[0000-0002-8279-9324]{O.S.~AbouZeid}$^\textrm{\scriptsize 40}$,    
\AtlasOrcid{N.L.~Abraham}$^\textrm{\scriptsize 156}$,    
\AtlasOrcid[0000-0001-5329-6640]{H.~Abramowicz}$^\textrm{\scriptsize 161}$
% ...
\bigskip
\\

% ...
$^{36}$CERN, Geneva; Switzerland.\\
% ...
$^{40}$Niels Bohr Institute, University of Copenhagen, Copenhagen; Denmark.\\
% ...
$^{53}$II. Physikalisches Institut, Georg-August-Universit\"{a}t G\"ottingen, G\"ottingen; Germany.\\
% ...
$^{94}$Department of Physics, Royal Holloway University of London, Egham; United Kingdom.\\
% ...
$^{102}$CPPM, Aix-Marseille Universit\'e, CNRS/IN2P3, Marseille; France.\\
$^{103}$Department of Physics, University of Massachusetts, Amherst MA; United States of America.\\
% ...
$^{128}$Homer L. Dodge Department of Physics and Astronomy, University of Oklahoma, Norman OK; United States of America.\\
% ...
$^{156}$Department of Physics and Astronomy, University of Sussex, Brighton; United Kingdom.\\
% ...
$^{161}$Raymond and Beverly Sackler School of Physics and Astronomy, Tel Aviv University, Tel Aviv; Israel.\\
% ...
$^{167}$Department of Physics, University of Toronto, Toronto ON; Canada.\\
% ...

\end{flushleft}

 
\end{document}